\documentclass[twocolumn,amsmath,amssymb,secnumarabic,prl]{revtex4}

\usepackage{graphicx,cancel}
\usepackage{ulem,latexsym}

\countdef\preprint=0 \countdef\tube=12
\ifnum\tube<10

\else

\fi

\begin{document}
\setlength{\mathindent}{0.2cm}

\title{Electron--electron correlation in graphite: a combined angle-resolved photoemission and first-principle study}

\author{A.~Gr\"{u}neis$^{1}$,
C.~Attaccalite$^{2}$, T.~Pichler$^{1}$, V.~Zabolotnyy$^{1}$,
H.~Shiozawa$^{1}$, S.L.~Molodtsov$^{3}$, D.~Inosov$^{1}$,
A.~Koitzsch$^{1}$, M.~Knupfer$^{1}$, J.~Schiessling$^{4}$,
R.~Follath$^{5}$, R.~Weber$^{5}$, P.~Rudolf$^{6}$, L.~Wirtz$^{2}$,
A.~Rubio$^{7}$}

\affiliation{$^{1}$IFW Dresden, P.O. Box 270116, D- 01171 Dresden,
Germany}

\affiliation{$^{2}$Institute for Electronics, Microelectronics,
and Nanotechnology, B.P. 60069, 59652 Villeneuve d'Ascq Cedex,
France}

\affiliation{$^{3}$ Institut f\"ur Festk\"orperphysik, TU Dresden,
Mommsenstrasse 13, D-01069 Dresden, Germany}

\affiliation{$^4$Uppsala University, Dept. of Physics, PO BOX 530
75 121, Uppsala, Sweden}

\affiliation{$^5$BESSY II, Albert-Einstein-Str. 15, 12489 Berlin,
Germany}

\affiliation{$^6$Materials Science Centre, Rijksuniversiteit
Groningen, Nijenborgh 4, NL-9747 AG, Groningen, The Netherlands}

\affiliation{$^7$Dept. Fisica de Materiales, Donostia International Physics Center, Spain\\
European Theoretical Spectroscopy Facility (ETSF), Spain}

\date{\today}
\begin{abstract}
The full three--dimensional dispersion of the $\pi$-bands, Fermi
velocities and effective masses are measured with angle--resolved
photoemission spectroscopy and compared to first-principles
calculations. The band structure by density-functional theory
underestimates the slope of the bands and the trigonal warping
effect. Including electron-electron correlation on the level of
the GW approximation, however, yields remarkable improvement in
the vicinity of the Fermi level. This demonstrates the breakdown
of the independent electron picture in semi-metallic graphite and
points towards a pronounced role of electron correlation for the
interpretation of transport experiments and double-resonant Raman
scattering for a wide range of carbon based materials.
\end{abstract}
\maketitle

\begin{figure*}
\begin{tabular}{c}
\vspace{-0.5cm}\hspace{-0.5cm}\includegraphics[width=3.59cm]{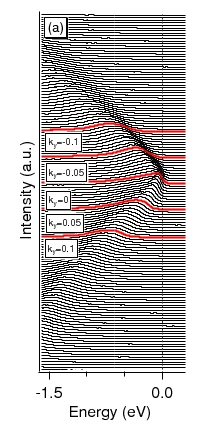}
\hspace{-0.5cm}\includegraphics[width=5cm]{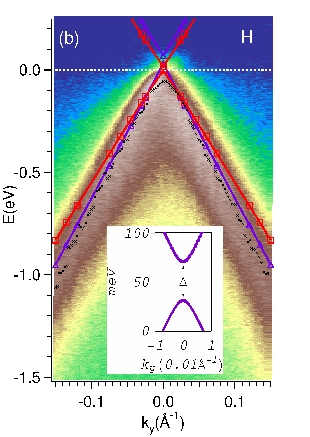}
\hspace{-0.5cm}\includegraphics[width=5cm]{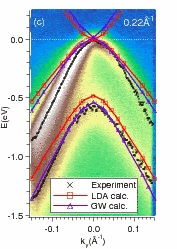}
\hspace{-0.5cm}\includegraphics[width=5cm]{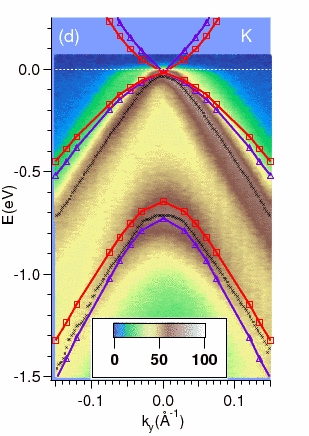}
\end{tabular}
\caption{(color online) (a) Raw EDC data for the cut through $H$. Cuts
at special $k_y$ (values in $\rm \AA^{-1}$) are indicated by red lines.
Photoemission intensity of cuts through the
3D BZ of graphite at (b) $H$, (c)
$\rm k_z=0.22\AA^{-1}$ and (d) $K$ point. Along with the
photoemission intensity, we show results of LDA ($\Box$, red
line) and GW ($\bigtriangleup$, blue line) calculations. The inset
in (b) shows a magnification of the TB-GW calculation for the small hole pocket and the gap $\Delta$. The photon energies used were (b) 100~eV, (c)
25~eV and (d) 83~eV. The intensity in (c) is multiplied by a factor of 2 in order to visualize the low intensity lower $\pi$ band. \label{fig:arpes1}}
\end{figure*}
 Recently graphene has been investigated
as a prototype system to address basic questions of quantum
mechanics~\cite{geim07-review,lanzara06-graphite,geim06-kleinparadox}
(relativistic Dirac fermions) as well as for high speed
semi--metal field effect transistors in emerging nanoelectronic
devices~\cite{novoselov06-graphite}. Many of these results are
based on its peculiar electronic properties, i.e. an isotropic and
linear dispersion close to the Fermi level ($E_F$). In low
dimensional and strongly anisotropic systems correlation effects
play a crucial role in understanding and describing the electronic
band structure. Kinks in the quasiparticle (QP) dispersions and
lifetimes were observed and interpreted as band renormalization
due to electron--phonon~\cite{lanzara06-annals_graphite} and
electron--plasmon~\cite{rotenberg06-graphite} interactions and
band structure effects~\cite{spataru01-graphite}. Its electronic
properties are also very sensitive to stacking and the number of
layers~\cite{henrard06-hpoint}. In bilayer graphene a gap that
could be tuned by doping was
observed~\cite{rotenberg06-graphite_bilayer}. For few layered
graphene, the parent compound, graphite, is the key to
understanding these new phenomena. Interlayer coupling in an $AB$
stacking sequence leads to the formation of electron and hole
pockets responsible for the semi--metallic character in graphite.
The linear dispersion is broken and only if we have an $AA$
stacking the linear dispersion remains. Nevertheless, at the $H$
point of graphite~\cite{lanzara06-graphite} the $\pi$-band
dispersion is close to linear and has been interpreted as
Dirac-Fermion like. Much less is known about the quantitative
description of electron--electron correlations in these graphitic
systems. Angle--resolved photoemission (ARPES) studies indicated
that local density approximation (LDA) gives a dispersion that is
too flat and a scaling has to be applied in order to fit the
experimental dispersion of few--layer graphene and graphite. For
the scaling, values of
$\sim$10\%~\cite{heske98-graphite,strocov01-graphite} and
$\sim$20\%~\cite{rotenberg06-preprint1,zhou05-graphite} have been
reported in the literature. Correlation effects can cause the
recently discovered quantitative shortcomings~\cite{Graf07} of the
double-resonant model~\cite{thomsen00d} for Raman scattering in
graphene and graphite. Furthermore, it is important to know the
exact $k_z$ dispersion, because it is responsible for the
conductivity perpendicular to the graphene layers.
\par
In this letter we report on a combined ARPES and theoretical
ab--initio QP study of the three dimensional $\pi$ band structure
and the Fermi surface in graphite single crystals. ARPES is best
for studying correlations since it probes the complex self--energy
function which contains the electronic interactions. We elucidate
the full electronic QP dispersion perpendicular to the layers and
show directly the importance of the effects of electronic
correlation and the influence on transport and resonance Raman in
these layered materials.
Experiments were done at BESSY~II using the UE112-PGM2 beamline
and a Scienta SES~100 analyzer yielding a total energy resolution
of 15~meV and a momentum resolution better than $\rm
0.01~\AA^{-1}$. Natural graphite single crystals with $AB$
stacking were cleaved in--situ to give mirror--like surfaces and
were measured within 12~h after cleavage in a vacuum better than
$10^{-10}$~mbar. The samples were mounted on a three axis
manipulator that was cooled down by liquid He to 25~K. The
three--step model for ARPES is employed for data
analysis~\cite{huefner-pes}. After a dipole transition from the
valence band to an unoccupied intermediate state, the electron
travels to the sample surface where it is scattered into a free
electron state outside the sample. When the electron leaves the
sample, its wavevector component in the direction perpendicular to
the sample surface is not conserved which makes 3D band mapping a
challenging task. We  assume that the intermediate state
dispersion is a parabola which is shifted downwards in energy with respect to the vacuum level by
the inner potential. We determine the inner potential by finding
the maximum splitting between the two valence $\pi$ bands, which
occurs at $\rm k_z=0$, i.e. $K$. This procedure yields an inner
potential of $V_0=16.4\pm 0.1$~eV and allows one to accurately
determine the full $k_z$ dispersion by changing the photon
energy~\cite{huefner-pes}. Deviations from a parabolic dispersion
of the intermediate state would change the $k_z$ assignment as
described in~\cite{strocov00-graphite,Krasovskii05-graphite,Krasovskii07-tite}.
However, as shown in detail below and in agreement with previous
experiments~\cite{lanzara06-graphite,rotenberg06-preprint1} for the
electronic states of graphite close to the Fermi level this error
is small.

\begin{figure}
\begin{minipage}[hc]{8cm}
\hspace{-0.0cm}\includegraphics[width=6.0cm]{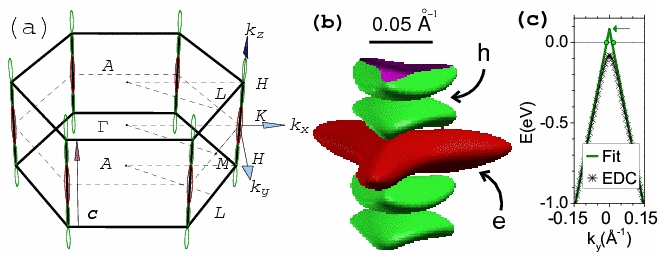}
\vspace{0.5cm}\\
\hspace*{-0.0cm}\includegraphics[width=7.5cm]{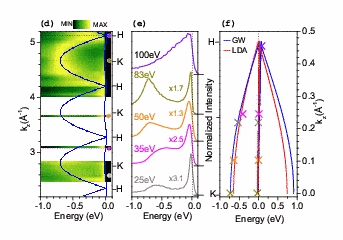}
\end{minipage}
\caption{(color online) (a) Graphite BZ with the high symmetry
points, reciprocal lattice vector ${\bf c}$ and the coordinate
system. The red (green) curves around $K$($H$) are the envelopes
of the measured electron(hole) pockets. (b) depicts the
interpolation of the 3D Fermi surface to experimental data. (c) Obtaining band maxima:
crosses denote symmetrized EDC data and the green line an
extrapolation. The hole pocket cross section with $E_F$ is denoted
by green circles. The arrow denotes the band maximum at $H$. (d)
photoemission intensity along $KH$ direction for photon energies
from 22~eV to 105~eV and GW calculations (blue line). The five
colored points along $KH$ denote $k_z$ for which we evaluate the
EDC maxima. (e) Raw EDCs for the $k_z$ denoted by colored points.
(f) The experimental EDC maxima and the calculated LDA and GW
dispersion.\label{fig:arpes2}}
\end{figure}

\begin{figure}
\begin{tabular}{cc}
\hspace{-3mm}\includegraphics[width=3.1cm]{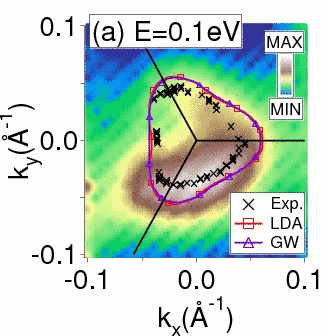}
\hspace{-2mm}\includegraphics[width=3.1cm]{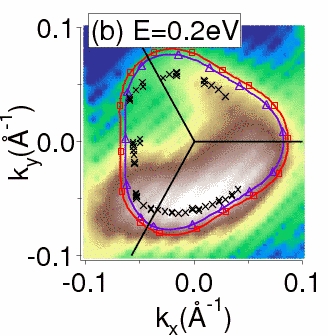}
\hspace{-2mm}\includegraphics[width=3.1cm]{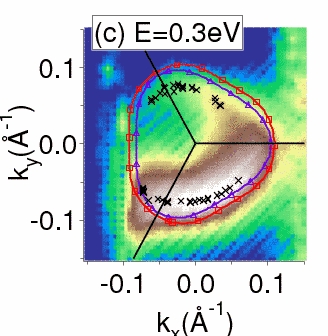}
\end{tabular}
\caption{(color online) Equi--energy contours of the photoemission
intensity (logarithmic scale) around the $KH$ axis for
$k_z=0.22$~$\rm \AA^{-1}$ corresponding to Fig.~1(c). Experimental
EDC maxima are denoted by crosses. Lines depict TB-GW (blue,
$\bigtriangleup$) and LDA (red, $\Box$) calculations,
respectively. Trigonal warping and the strong asymmetry in the
intensity can clearly be seen.}\label{fig:arpes3}
\end{figure}

The calculations of the electronic dispersion are performed on two
levels. First, we calculate the Kohn-Sham band-structure within
the LDA to density-functional theory (DFT)~\cite{abinit}.
Wave-functions are expanded in plane waves with an energy cutoff
at 25 Ha. Core electrons are accounted for by Trouiller-Martins
pseudopotentials. In the second step, we use the GW-approximation
\cite{hybert86,hedin65-gw,louie-gw} to calculate the self--energy
corrections to the LDA dispersion~\cite{self}. For the calculation
of the dielectric function $\epsilon(\omega,q)$ we use a
15$\times$15$\times$5 Monkhorst-Pack $k$ sampling of the first BZ,
and conduction band states with energies up to 100~eV above the
valence band (80~bands). The calculations are converged to 10~meV.
Since the GW calculations are computationally expensive a tight--binding fit to GW (TB-GW) was
performed in order to rapidly calculate the QP bands.

We now turn to the comparision of the ARPES data to the calculated
band structures. In Fig.~\ref{fig:arpes1} we show raw EDC data and
cuts of the $\pi$ bands that intersect the corners of the BZ at
$k_z=0.46$~$\rm\AA^{-1}$ (very close to $H$ point), at
$k_z=0.22$~$\rm\AA^{-1}$ and at $k_z=0$ ($K$ point). The cuts are
done along the $k_y$ axis [see coordinate system in
Fig.~\ref{fig:arpes2}(a)]. It is clear that for the cut through
the $H$ point, the bands are almost linear and only one $\pi$
valence band can be seen, whereas at the $K$ point, the degeneracy
is lifted and the bands are parabolic. The energy separation
between the two valence $\pi$ bands is getting smaller when moving
away from $K$ in $k_z$ direction. The LDA and GW calculation data
along the measured cut are also shown in Fig.~\ref{fig:arpes1}.
LDA  underestimates the slope of the bands while the self--energy
corrections from the GW calculation yield a better agreement. In
the inset of Fig.~\ref{fig:arpes1}(b) we show the gap $\Delta$ at
$H$ obtained by GW. It is clear that the dispersion is not linear
close to $H$ and the concept of Dirac Fermions breaks down. This
is important for transport properties, however for estimation of
the band maxima a linear extrapolation above $E_F$ is a good
approximation due to the extremely high curvature at the maximum
of the hole like band at $H$.

We can obtain the Fermi surface of the hole like band with a band
maximum above $E_F$ by the following method: we fit a linear
dispersion to the EDC maxima in a region below $E_F$ and then
extrapolate it to find the crossing of the extrapolated band with
the Fermi level. A dispersion very close to linear is predicted by
calculation [inset of Fig.~\ref{fig:arpes1}(b)] as the functional
form of the band at $H$. From the crossing of the linear
dispersion with $E_F$, we can then obtain an estimate for the
cross section of the hole pocket at $H$. Such a procedure is
needed in order to determine the maximum of the sparsely populated
band and has been used in the
literature~\cite{lanzara06-graphite}. For points with $k_z<0.5c$,
we choose a quadratic fit as the functional form around the band
maximum ($c$ is the length of the reciprocal lattice vector in $k_z$ direction). Assuming electron--hole symmetry for the sparsely
occupied band, we can estimate the cross sections with $E_F$. From
this data and some higher energy points the Fermi surface could be
interpolated with the standard tight--binding
Hamiltonian~\cite{dresselhaus81}.

The pockets we obtained are shown in Fig.~\ref{fig:arpes2}(a)
along with a sketch of the BZ and the coordinate system and reciprocal lattice vector $\bf c$. A detailed magnification of the
tiny electron and hole pockets is shown in
Fig.~\ref{fig:arpes2}(b) with the aspect ratio enlarged by a
factor 30 for easier viewing. The surfaces around $K$ and $H$
points are the electron and hole pockets, respectively. The
volumes inside the Fermi surface based on experiment 
yield carrier densities of $n_e=2.3\times 10^{19}$~$\rm cm^{-3}$
electrons and $n_h=1.8\times 10^{19}$~$\rm~cm^{-3}$ holes. In
Fig.~\ref{fig:arpes2}(c) we illustrate the procedure to obtain the
$\pi$ band maxima at $H$ which is located above $E_F$. In Fig.~\ref{fig:arpes2}(d) we show the photoemission intensity
along the $KH$ direction for photon energies between 22~eV and
105~eV. In Fig.~\ref{fig:arpes2}(e) the raw EDC data for selected
cuts along $KH$ are depicted and in Fig.~\ref{fig:arpes2}(f) the
EDC band maxima (folded back to the first BZ) are compared to
theory. The experimental results and theory have the same trend,
which is that the lower $\pi$ valence band comes close to the
upper one when moving from $K$ to $H$ point. The experimental
splitting between the $\pi$ valence bands is generally better
described by the GW approximation than by LDA.

In Fig.~\ref{fig:arpes3} the trigonal warping effect is
investigated. In order to accurately compare to calculations we
evaluate the maxima of the EDCs for equi--energy contours in the
$k_x, k_y$ plane. We have chosen $k_z=0.22$~$\rm \AA^{-1}$, which
lies approximately halfway between $K$ and $H$. The experimental
and calculated GW and LDA contours for three energies
0.1~eV~--~0.3~eV are shown in (a)--(c). LDA not only underestimates
the slope of the bands but also a simple scaling does not work
because a different scaling would be needed for $\Gamma$ and $M$
directions. Although the GW approximation does not describe the
experimental data in all directions, it is always closer to the
experiment than the LDA. Interestingly, it is in perfect agreement
with the experiment at $H$ point while the agreement is worse when
going away from $H$. This discrepancy might be due to the GW
approximation and resolved by inclusion of higher orders in the
calculation of the self--energy correction in a future work.

A strong asymmetry around the $KH$ axis in energy (trigonal
warping effect) and photoemission intensity is observed. The
trigonal warping effect of the ARPES data is underestimated by the
LDA and again better described in the GW approximation. Compared
to graphene we find that trigonal warping is higher by a factor of
$\sim 1.5$ when considering the distances in the two
high--symmetry directions for an equi--energy contour of 0.1~eV in
Fig.~\ref{fig:arpes3}. The strong asymmetry in the photoemission
intensity seen in this figure is attributed to the dipole matrix
element for the transition from a valence band state to an
unoccupied state 25~eV above $E_F$~\cite{shirley95-dipole}.

We now turn to a quantitative assessment of electron correlation
effects and their implications for Raman and transport studies of
graphite and related materials. The experimental and calculated
values for the energy separation of valence bands at $K$, the
gap $\Delta$ at $H$ and the Fermi velocity $v_F$ are listed
in Table~\ref{tab:fermivel}. We obtain the Fermi velocities of the linear band in $HA$ and $HL$ direction in a region up to 1.5~eV below $E_F$.
Since the trigonal warping effect has a minimum in this $k_z$ plane, the differences in these two directions are small.
The LDA values for $v_F$ are far off
and the differences of $v_F$ in $HA$ and $HL$ directions (trigonal
warping effect) are strongly overestimated.  On the other hand the
GW is in excellent agreement with the experiment.
\begin{table}
\begin{tabular}{c|c|c|p{2.5cm}|c}
  \hline
\hline
  Method & $v_F$ (HA) & $v_F$ (HL) &   energy \linebreak {separation} (eV) & $\Delta$ (eV) \\
\hline
ARPES& $1.06\pm 0.1$ & $1.05~\pm~0.1$ & $0.71~\pm~0.015$ & -- \\
GW & 0.99 & 1.02 & 0.71 & 0.037 \\
LDA & 0.77 & 0.87 & 0.62 & 0.022\\
\hline \hline
\end{tabular}
\caption{Fermi velocities at $H$ (in $10^{6}$ m/s), valence band
energy separation at $K$ and $\Delta$ obtained by experiment and
from calculations. }\label{tab:fermivel}
\end{table}

The dispersion of $D$ and $G'$ double resonant Raman (DRR)
bands~\cite{thomsen00d} with laser energy is proportional to
$v_{Ph}$/$v_F$, where $v_{Ph}$ is the slope of the transversal
optical phonon branch going through $K$ ($H$). The electronic
dispersion in $k_z$ broadens the DRR dispersion and with the
present data DRR spectra for 3D graphite can be calculated taking
into account the full electronic dispersion. For a given phonon
dispersion we thus expect the present data to reduce the slope of
the calculated DRR bands since the $v_F$ from our experiment are
blue--shifted with respect to the LDA calculation.

In addition, the gap $\Delta$ shows a large difference in
values with a GW value that is five times the TB parameter that
was originally fit to a magnetoreflectance
experiment~\cite{dresselhaus81}. Such a deviation can be
attributed to stacking faults such as $AA$ stacking, which has
$\Delta=0$. For a sample consisting of $AB$ stacking with many
stacking faults the effectively measured $\Delta$ would be lower
explaining the experimental value of $\Delta=0.008$~eV in
pyrolytic graphite~\cite{dresselhaus81}. A finite value of
$\Delta$ causes a breakdown of the linear bands at the $H$ point.
Their shape becomes parabolic with a very large curvature and thus
a small absolute value of the effective mass.

From the 3D Fermi surface the transport properties of graphite can
be calculated from a Drude model where the carrier densities and
the effective masses can be directly taken from the ARPES
experiment. For the $k_z$ averaged electron and hole masses, we obtain
$m_e^*=0.10m_0$ and $m_h^*=-0.04m_0$, respectively that is in good
agreement with a previous magnetoreflectance study ($m_0$ is the
free electron mass)~\cite{mendez79-effective_massa}. For the
scattering time $\tau$ we use a value of $\tau\sim 200$~fs from
pump--probe experiments~\cite{hertel01-fastoptic}. We get a DC
conductivity $\sigma=n_ee^2\tau/m_e^*+n_he^2\tau/m_h^*$ of
$\sigma=\rm 3.9\times 10^4~\Omega^{-1}cm^{-1}$. The values agree
nicely with transport measurements that yield $\rm 2.5\times
10^5~\Omega^{-1}cm^{-1}$~\cite{morelli84-resistivity} and $\rm
2.5\times 10^4~\Omega^{-1}cm^{-1}$~\cite{dresselhaus81} when
considering that the literature values differ by an order of
magnitude.

In conclusion we have performed ARPES of graphite and compared the
measured QP dispersion to ab-initio calculations. We have found
that the band dispersions are better described by the GW
approximation, however around the $K$ point that agreement becomes
poorer and many-body approximation schemes beyond GW may improve
this result. Moreover, for this system the concept of Dirac
Fermions breaks down. This highlights the importance of electron
correlations which renormalize the electronic dispersion resulting
in a blue shift of the gap at $H$ and an increased $v_F$. A direct
comparison of a scaled LDA QP dispersion to transport and Raman
experiments is not possible due to an anisotropy of the scaling
needed for the LDA around the $KH$ axis. TB-GW allowed us rapid calculation of the
QP bands.

A.G. acknowledges a Marie Curie Individual Fellowship (COMTRANS)
from the European Union. T.P. acknowledges DFG projects PI 440/3
and 440/4. C.A. and L.W. acknowledge support from the French
national research agency. A.R. is supported by the EC Network of
Excellence Nanoquanta (NMP4-CT-2004-500198), SANES project
(NMP4-CT-2006-017310), Basque Country University (SGIker Arina),
MCyT. Calculations were performed at Barcelona supercomputing
center and at Idris (Paris). We thank Andrea Marini for making his
GW-code {\tt SELF} available to us.

\end{document}